%% file: main.tex
\documentclass[acmtog]{acmart}
\acmSubmissionID{1315}
\usepackage[table]{xcolor}
\usepackage{macros}
\usepackage{booktabs}
\usepackage{makecell}
\usepackage{wrapfig,calc,placeins,microtype}
\usepackage{mathtools}
\usepackage[noabbrev,capitalise]{cleveref}
\usepackage{enumitem}
\usepackage{hyperref}
\usepackage{amsthm}
\usepackage{soul}
\usepackage{float}

\crefname{figure}{Figure}{Figures}
\Crefname{figure}{Figure}{Figures}

\usepackage[ruled]{algorithm2e}

\SetAlFnt{\small}
\SetAlCapFnt{\small}
\SetAlCapNameFnt{\small}
\SetAlCapHSkip{0pt}

\newcommand{\ie}{\emph{i.e.}\ }

\acmJournal{TOG}
\acmVolume{44}
\acmNumber{6}
\acmArticle{1} % TODO update this when info is sent out
\acmYear{2025}
\acmMonth{12}
\acmPrice{}
\acmDOI{10.1145/3763298}
\setcopyright{cc}
\setcctype{by}
\begin{document}
% Title portion
\title{PoissonNet: A Local-Global Approach for Learning on Surfaces}

\author{Arman Maesumi}
\email{arman.maesumi@gmail.com}
\orcid{0000-0001-7898-8061}
\affiliation{%
  \institution{Brown University}
  \country{USA}
}

\author{Tanish Makadia}
\email{tanish_makadia@brown.edu}
\orcid{}
\affiliation{%
  \institution{Brown University}
  \country{USA}
}

\author{Thibault Groueix}
\email{thibault.groueix.2012@polytechnique.org}
\orcid{0000-0002-7984-8252}
\affiliation{%
  \institution{Adobe Research}
  \country{USA}
}

\author{Vladimir G. Kim}
\email{vova.g.kim@gmail.com}
\orcid{0000-0002-3996-6588}
\affiliation{%
  \institution{Adobe Research}
  \country{USA}
}

\author{Daniel Ritchie}
\email{daniel_ritchie@brown.edu}
\orcid{0000-0002-8253-0069}
\affiliation{%
  \institution{Brown University}
  \country{USA}
}

\author{Noam Aigerman}
\email{noamaigerman@gmail.com}
\orcid{0000-0002-9116-4662}
\affiliation{%
  \institution{Universit\'e de Montr\'eal}
  \country{CA}
}

\renewcommand\shortauthors{Maesumi, A. et al}

\begin{abstract}
\input{sec_abstract}
\end{abstract}



\begin{teaserfigure}
  \includegraphics[width=\textwidth, trim=0 0.25cm 0 0]{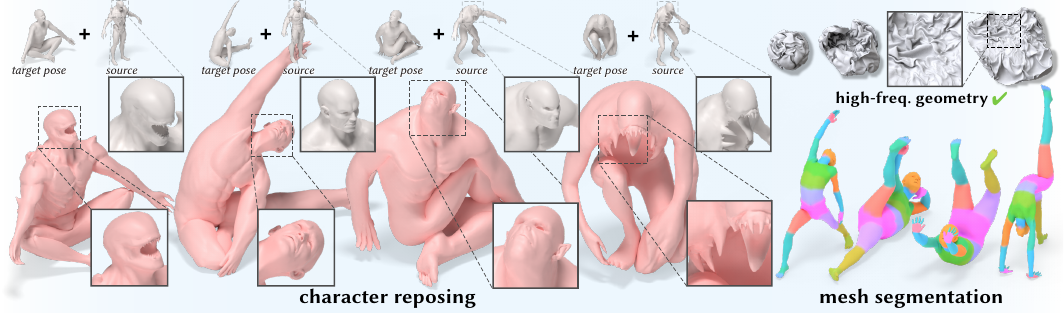}
  \caption{We develop a general neural architecture for learning on surfaces that uses a local-global construction, resulting in a framework that is highly accurate for processing detailed meshes while being more efficient than comparable methods. \emph{Left:} We train PoissonNet on source-to-target shape deformation for humanoid characters. Our model is able to generalize to in-the-wild geometries while preserving fine details and obeying the provided pose parameters (see reference \emph{target pose} exemplars). \emph{Bottom right:} Our method is general and can be applied broadly to learning tasks on surfaces---e.g. for finely segmenting human bodies. \emph{Top right:} PoissonNet is able to represent extremely high-frequency geometry, such as a crumpling paper ball with 300k faces.}
  \label{fig:teaser}
\end{teaserfigure}

\maketitle

\input{sec_intro2}
\input{sec_related_works}
\input{sec_method}
\input{sec_implementation}
\input{sec_results}
\input{sec_conclusion}

\begin{acks}
This material is based upon work supported by: National Science Foundation Graduate Research Fellowship under Grant No. 2040433; NSERC Discovery grant RGPIN-2024-04605, ``Practical Neural Geometry Processing'';
FRQNT Établissement de la relève professorale 365040, ``Calcul rapide et léger des déformations à l'aide de réseaux neuronaux''; and a gift from Adobe. Part of this work was done while Arman Maesumi was an intern at Adobe Research. The authors thank Qingnan Zhou for providing preprocessed Thingi10k data, as well as Nick Sharp, Alec Jacobson, and Derek Liu for fruitful discussions.
\end{acks}
\bibliographystyle{ACM-Reference-Format}
\bibliography{bib}
\clearpage
\appendix
\input{supp}
\end{document}

%% file: sec_abstract.tex
Many network architectures exist for learning on meshes, yet their constructions entail delicate trade‑offs between difficulty learning high-frequency features, insufficient receptive field, sensitivity to discretization, and inefficient computational overhead. Drawing from classic local-global approaches in mesh processing, we introduce PoissonNet, a novel neural architecture that overcomes all of these deficiencies by formulating a local-global learning scheme, which uses Poisson's equation as the primary mechanism for feature propagation. Our core network block is simple; we apply learned \emph{local} feature transformations in the gradient domain of the mesh, then solve a Poisson system to propagate scalar feature updates across the surface \emph{globally}. Our local‑global learning framework preserves the features's full frequency spectrum and provides a truly global receptive field, while remaining agnostic to mesh triangulation. Our construction is efficient, requiring far less compute overhead than comparable methods, which enables scalability---both in the size of our datasets, and the size of individual training samples. These qualities are validated on various experiments where, compared to previous intrinsic architectures, we attain state-of-the-art performance on semantic segmentation and parameterizing highly-detailed animated surfaces. Finally, as a central application of PoissonNet, we show its ability to learn deformations, significantly outperforming state-of-the-art architectures that learn on surfaces.
\href{https://github.com/ArmanMaesumi/poissonnet}{\color{blue}https://github.com/ArmanMaesumi/poissonnet}

%% file: sec_intro2.tex
\input{tables/method_comparison}

\section{Introduction}
Recent years have seen an explosion in techniques for deep learning on surfaces represented as triangle meshes. As opposed to point clouds and voxel grids, meshes are a representation that can easily encode highly-detailed geometry, provide an appropriate discretization of a Riemannian manifold, encode explicit topological information, and enable computations on the underlying surface (e.g. via finite elements). For these reasons, meshes remain the primary choice of representation for a wide array of applications in graphics.

Current state-of-the-art learning methods for meshes follow an \emph{intrinsic} approach by employing the surface's differential operators (e.g., the Laplacian, gradient, curl, divergence). At each block of the neural network, the differential operators are used to transform intermediate feature fields (e.g. by taking the divergence of a vector field of features~\cite{deltaconv}); alternatively, the differential operators may define a partial differential equation (PDE) whose solution serves as the transformed signal~\cite{diffusionnet, vectorHeatNetwork}. This approach provides the means to treat the surface with proper tools from differential geometry, and endows these methods with crucial properties, such as \emph{triangulation agnosticism} (different discretizations of the same shape lead to similar outputs). 

There are many ways to incorporate differential operators in a learning framework, often resulting in intricate constructions, which in turn exhibit particular deficiencies that reoccur across all existing methods, e.g.,
a limited receptive field due to locality of the chosen operators; inability to represent features in their full frequency spectrum; sensitivity to surface discretization; and, expensive computation and memory footprints.

In this paper, we devise a simple and straightforward intrinsic learning approach that overcomes the above issues. We achieve this by formulating our network through \emph{Poisson's equation} --- a particular PDE, which we argue is a natural choice for this task. Poisson's equation is one of the most ubiquitous PDEs in graphics, appearing in cornerstone algorithms such as As-Rigid-As-Possible~\cite{ARAP_modeling_sorkine_2007}, Poisson Surface Reconstruction~\cite{poissonrecon_kazhdan_2006}, in image editing~\cite{fattal2002,poissonImageEditing}, and recently as a proxy to learning deformations' gradients~\cite{NJF}. Surprisingly, no work has leveraged Poisson's equation for feature learning on meshes; rather, its use has been limited to end-stage ``integration'' steps common in aforementioned algorithms.

Intuitively, Poisson's equation acts as a bridge between the gradients of signals, and the signals themselves: if the gradient operator transforms signals into their spatial gradients, then Poisson's equation can be seen as its inverse---conceptually akin to an integration operator (see Section \ref{sec:prelims} for further discussion). 

Using this fact, we design a network architecture that alternates between the gradient domain and the functional domain, similar to classic local-global algorithms in graphics. Concretely, each block of our network takes the gradient of the signal (i.e. features), applies \textit{local} learned transformations in the gradient domain, and then solves Poisson's equation to obtain \textit{global} (i.e. not localized in their receptive field) feature updates in the scalar domain, thereby placing gradients as \emph{first class citizens} in our framework. 

Operating primarily in the gradient domain is a crucial property for learning over meshes. Indeed, throughout the literature, we observe that previous intrinsic learning methods consistently identify the gradient operator as a critical component of their architectures, with some noting the framework significantly underperforms without gradient features  (e.g. Figure 6 in Sharp et al.~\shortcite{diffusionnet}). 

Whereas prior methods for learning on surfaces trade off along several key properties (see Table \ref{tab:methodComparison}), our method yields the first network that satisfies them simultaneously:
\begin{enumerate}
    \item \textbf{Full spectrum.} Our network features retain their native frequency components without any spectral truncation, preserving high-frequency details while avoiding expensive precomputation of eigenbases.
    \item \textbf{Global receptive field.} As an integral‐like operator, our proposed network block efficiently propagates local feature updates across the entire surface.
    \item \textbf{Triangulation agnosticism.}
    The core mechanism in our network approximates a well-defined object: the continuous Poisson's equation. This allows PoissonNet to produce near-identical predictions under changes in mesh discretization (subdivision, simplification, remeshing, corruption, etc.).
    \item \textbf{Efficient computational footprint.} Our method is scalable: PoissonNet can operate on high-resolution meshes and forgo lengthy pre-computation before training and inference, which facilitates training on large datasets.
\end{enumerate}
We empirically validate these properties on a range of canonical applications, such as shape segmentation, deformation, and learning high-frequency signals on surfaces. In all experiments, PoissonNet achieves state-of-the-art performance, while remaining far more efficient than previous methods with comparable capabilities. Code is available at: \href{https://github.com/ArmanMaesumi/poissonnet}{\color{blue}https://github.com/ArmanMaesumi/poissonnet}

%% file: tables/method_comparison.tex
\begin{table*}[ht]
\centering
\caption{A bird’s‑eye view of trade‑offs associated with various methods for learning on surfaces. Columns correspond to: \textbf{Full Spectrum}: whether features are propagated in their full frequency spectrum, or spectrally truncated; \textbf{Spatial Support}: the effective receptive field of each atomic block in the network; \textbf{Triangulation Agnostic}: changes in triangulation produce near-identical outputs; \textbf{Precompute}: amount and type of required per‐mesh precomputation; \textbf{Inference}: per‐sample inference latency (ignoring precompute); and \textbf{Scalable}: ability to scale up training or model size -- affected by amount of precompute (dataset-bound) and inference efficiency (mesh-bound). Green entries indicate desirable extremes, red entries indicate undesirable extremes (spectral truncation, expensive per‐mesh precomputation, slow runtime), with intermediate hues reflecting partial trade‑offs. \textbf{PoissonNet} is the first method to simultaneously encompass: feature propagation in the full eigenspectrum, global spatial support (via a sparse Poisson system that is \emph{efficiently} solvable across our network), agnosticism to mesh discretization, and ability to scale in the number of data samples and mesh size, thereby addressing the fundamental limitations of prior intrinsic architectures. \textsuperscript{\textdagger}DeltaConv is a point-based method, though it operates using similar constructs as the other methods in this table. Its entry
\emph{``No''} under \emph{Triangulation Agnostic} reflects lack of discretization invariance: local K‑NN neighborhoods (and thus the layer's behavior) change with sampling density. $^\ddagger$ \emph{hom.} and \emph{inhom.} indicate that DiffusionNet's heat equation is homogeneous, whereas Poisson’s equation is inhomogeneous (see Section \ref{sec:relatedWork} for discussion).}
\vspace{-1em}
\rowcolors{2}{seabornblue!25}{seabornblue!75}
\begin{tabular}{l c c c c c c}
\toprule
\thead[l]{Method}
&\thead{Full\\Spectrum}
&\thead{Spatial\\Support$^\ddagger$}
&\thead{Triangulation\\Agnostic}
&\thead{Precompute}
&\thead{Inference}
&\thead{Scalable}\\
\midrule
\textbf{PoissonNet (ours)}  & \good{Yes} & \good{Global \emph{(inhom.)}} 
 & \good{Yes} & \good{Single factorization} & \good{Fast} & \good{Yes}\\

\textbf{DiffusionNet (direct)} \shortcite{diffusionnet} & \good{Yes} & \okay{Learned \emph{(hom.)}} & \good{Yes} & \bad{Many factorizations} & \bad{Slow} & \bad{Mesh-bound}\\

\textbf{DiffusionNet (spectral)} \shortcite{diffusionnet} & \bad{Truncated} & \okay{Learned \emph{(hom.)}}& \good{Yes} & \bad{Eigenbases} & \good{Fast}  & \badish{Dataset-bound}\\

\textbf{DeltaConv}\textsuperscript{\textdagger} \shortcite{deltaconv} & \good{Yes} & \badish{Local + pooling}& \bad{No} & \good{K-NN} & \good{Fast}  & \good{Yes}\\

\textbf{HodgeNet} \shortcite{hodgenet_smirnov_2021} & \bad{Truncated} & \bad{Local} & \good{Yes} & \bad{Eigenbases} & \bad{Slow}  & \badish{Dataset-bound}\\

\textbf{Harmonic Surface Net} \shortcite{HSN_wiersma_2020} & \good{Yes} & \bad{Local} & \bad{No} & \bad{Parallel transport} & \bad{Slow}  & \bad{Mesh-bound}\\

\textbf{MeshCNN} \shortcite{meshcnn_hanocka_2019} & \good{Yes} & \badish{Local + pooling} & \bad{No} & \good{N/A} & \bad{Slow}  & \bad{Mesh-bound}\\

\bottomrule
\end{tabular}
% \vspace{-1em}
\label{tab:methodComparison}
\end{table*}

%% file: sec_related_works.tex
\begin{figure*}[!t]
    \centering
    \includegraphics[width=\linewidth, trim=0 0.2cm 0 0]{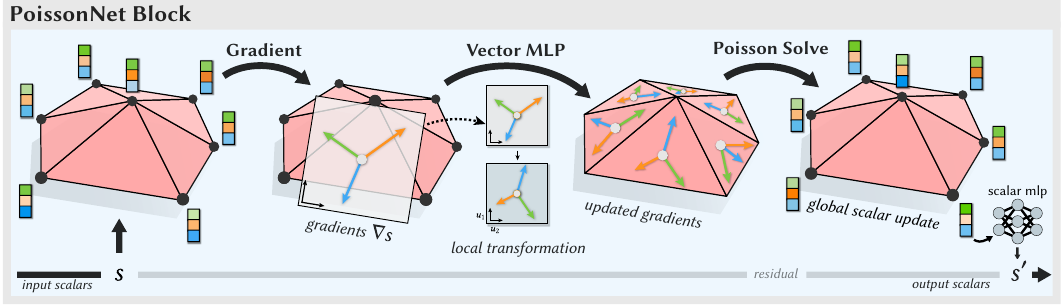}
    \caption{We illustrate our \emph{PoissonNet block} on a surface patch of a triangle mesh. Our block begins by computing spatial gradients of the incoming scalar features---for demonstrative purposes we depict the signal as having three channels (shown as green, orange, blue) with color indicating signal intensity. The gradient features are transformed locally in each tangent basis (denoted by basis vectors $u_1,u_2$) via a Vector MLP, which induces a linear combination of rotated and scaled gradients on each face. We then solve a global Poisson system using the transformed vector fields, producing new scalar features on vertices that are updated locally using a scalar (per-vertex) MLP. This process is repeated for $N$ blocks, thereby producing a final feature representation on the shape.}
    \label{fig:pipeline}
\end{figure*}
\section{Related Work}
\label{sec:relatedWork}

\paragraph{Learning on surfaces.} The first works to apply deep learning to surfaces considered point clouds sampled  from the surface, starting with the seminal PointNet~\cite{pointnet_qi_2017}, later extended to use a local receptive field~\cite{pointnet++_qi_2017, dgcnn_wang_2019, pointnext_koyejo_2022} and attention mechanisms~\cite{pointtransformer_Zhao_2021, pointvoxel_transformer_Zhang_2022, pointtransformev2_Koyejo_2022, pointtransformev3_Wu_2024_CVPR, yu2022point};  see~\cite{pointcloud_survey_guo_2020} for a survey. However, without knowledge of connectivity, point clouds cannot encode highly-detailed geometry, and the network may get confused by nearby points representing geodesically distant regions. 

To leverage the surface's connectivity, several approaches aim to generalize the notion of a convolution, either by flattening local patches to the plane~\cite{boscaini2016learning, fey2018splinecnn, monti2017geometric,bronstein2017geometric, dynamic_simonovsky_2017, geodesicConvo_masci_2015}, or via notions of equivariance~\cite{de2020gauge, he2020curvanet, fieldconvo_mitchel_2021, mdgcnn_poulenard_2018, HSN_wiersma_2020, cgconv_Yang_2021, sun2020zernet}. Others treat the mesh as a graph~\cite{dynamic_simonovsky_2017}, or apply a Recurrent Neural Network with  random walks on the mesh~\cite{meshwalker_lahav_2020}. MeshCNN~\cite{meshcnn_hanocka_2019} learns task-specific pooling strategies along with edge collapses. 
In contrast to our method, these methods are not \textit{triangulation agnostic}, i.e., changes in surface triangulation produce drastically different outputs, causing such networks to 1) learn spurious features that do not generalize to out-of-distribution geometries, and 2) mistakenly couple the mesh's resolution with the network's receptive field. We note that, in this context, triangulation agnosticism does not imply that a given architecture is invariant to all discretizations of an underlying surface; e.g. our method is still subject to discretization error.

\paragraph{Intrinsic learning on meshes.}
In order to perform triangulation-agnostic learning on surfaces, many works turn to differential operators derived from the meshes themselves, whose constructions are guaranteed to be triangulation agnostic. One popular approach is to leverage the spectral domain, performing a Fourier-like transform using the eigenmodes of the Laplace-Beltrami operator. Several methods leverage the Functional Maps Framework~\cite{ovsjanikov2012functional} in a deep learning setting~\cite{Litany_2017_ICCV,Yi_2017_CVPR, halimi2019unsupervised, roufosse2019unsupervised, donati2020deep, attaiki2021dpfm}. HodgeNet~\cite{hodgenet_smirnov_2021} extends spectral learning to vector fields and area forms. However, the spectral basis is represented as a dense matrix, which has a large memory footprint and is slow to compute (see Table \ref{tab:methodComparison}). Hence, these methods operate in the low-frequency part of the spectrum (the first $k$ eigenfunctions), hindering their ability to represent high-frequency signals. By contrast, PoissonNet does not rely on a spectral basis, and hence can capture the full frequency spectrum of signals.

DiffusionNet~\cite{diffusionnet} stands as the work closest to ours. Similar to our method, it propagates signals over surfaces by solving a PDE: the homogeneous heat equation. This PDE is often approximated via a single implicit integration step. In a learning framework, however, this becomes highly inefficient, as each learned diffusion time induces a distinct linear system, each needing a factorization that cannot be precomputed. As such, DiffusionNet instead solves its PDE in the spectral domain, restricting to the lower frequency range, which, as explained above, leads to loss of expressivity (see Fig.~\ref{fig:crumpleComparison}). Since the heat equation acts only as a radially-symmetric filter (see Fig. 6 in their paper), DiffusionNet uses the gradient operator to re-introduce directionality into their filters post-factum. Moreover, the homogeneous heat equation converges to a constant global average at steady state, erasing all structure when propagating features globally. By contrast, we directly solve Poisson's equation, which by construction: offers nontrivial global feature couplings (rather than producing constant signals); incorporates directional features directly into the PDE; and forgoes spectral bases entirely. These properties translate to superior performance and efficiency in several experiments, as discussed in Section \ref{sec:results}.

DeltaConv~\cite{deltaconv} stands as another close work, as it defines convolutions on surfaces by combining \textit{local} differential operators. However, since the operators are local, these convolutional layers have a local receptive field, which geometrically shrinks as the mesh is refined---requiring a deeper network for propagation of signals across the mesh, implying the method is not agnostic to sampling density. This is in contrast to our method, in which each layer has a global receptive field that approximates a well-defined continuous operation; \ie, different triangulations of the same underlying geometry produce nearly identical outputs, leading to triangulation agnosticism. Additionally, DeltaConv is published as a point-based method, \ie, their operators are defined through K-nearest neighbors, yielding artifacts in our experiments.

Neural Jacobian Fields (NJF)~\cite{NJF} also stands as inspiration for our method---though it is not a general learning architecture, but rather a method for learning \textit{deformations} of meshes. NJF trains a standard MLP to predict deformation gradients (Jacobians) as a neural field. Poisson's equation is then used in the final layer to produce a mapping from the deformation's gradient. In NJF, Poisson's equation is not a means to propagate learned features over meshes. By contrast, in our method, each consecutive network block solves Poisson's equation to globally propagate learned features over the domain. Additionally, NJF predicts an \textit{extrinsic} signal (i.e. not defined in a local coordinate frame), which is then projected into the mesh's tangent space. This is in contrast to our intrinsic transformations of gradients in the mesh's tangent space, which is critical for learning tasks, as it makes each network block rotation-invariant, leading to more efficient learning and robustness.

\paragraph{Learning Mesh Deformations.} 
As a primary benchmark and application for our network, we show that PoissonNet can serve as a strong backbone for learning mesh deformations. Mesh deformation is a long-standing problem in geometry processing with applications in animation~\cite{sumner2004deformation}, registration~\cite{bogo2014faust}, and geometric modeling~\cite{gao2019sdm}. 

Gradient-domain computation commonly appears in this scenario, using maps' Jacobians for surface parameterization~\cite{Levy02,Liu08,Rabinovich17,Smith15,Xingyi20,Shuller13,Aigerman13,Kovalsky14,Lipman12,Myles13,Weber14,Li18} or for deformation~\cite{Lipman04,Sorkine04,sumner2004deformation,Yu04}.

Many other ways to parameterize deformations as \emph{rigs} have been developed over the years~\cite{skinningcourse:2014,Fulton:LSD:2018,jacobson2011bounded,kavan2008geometric,lipman2008green,ju2005mean}, which have been used in a deformation-learning setting, e.g., using skeletons~\cite{AnimSkelVolNet,RigNet,Holden:inverse_rig:2015,li2021learning,liu2025riganything}, handles~\cite{liu2021deepmetahandles}, or cages~\cite{wang2019neural,sun2024tuttenet}. Some methods combine rig-driven deformation with non-linear residuals per discrete element
~\cite{Bailey:2018:FDD,bailey2020fast,Romero:2021,Zheng:secondary_motion:2021,yin2021_3DStyleNet}. 

Alternatively, for given template meshes~\cite{bogo2014faust,varol17_surreal,SMAL:2017,STAR:2020} one can directly predict a fixed-size tensor of vertex coordinates~\cite{anguelov2005scape,Bogo:ECCV:2016,Shen:2021}, possibly by directly learning the deformations' gradients~\cite{tan2018meshvae,gaovcgan2018}.  

We evaluate PoissonNet as a backbone, used together with a final layer provided by NJF~\cite{NJF}, discussed above, as it provides a triangulation-agnostic method for predicting deformations, which has since proven highly effective in scenarios such as temporal sequences~\cite{temporalNJF}, face rigging ~\cite{qin2023NFR}, and text/image-driven generative deformation~\cite{Gao23,Kim25,Yoo24}.

%% file: sec_method.tex
\section{Preliminaries}
\label{sec:prelims}
\paragraph{Tangent spaces and local coordinates.}We consider meshes with vertices $\V$ and triangles $\F$. Each triangle $\tri\in\F$ defines its own \textit{tangent space}, denoted $T_{\tri}$ --- a 2-dimensional linear subspace of $\Reals^3$, consisting of all vectors tangent to $\tri$. 
We choose an (arbitrary) orthonormal basis $\B_\tri=\{U_1,U_2\},\ U_i\in\Reals^3$ which serves as the tangent space's \textit{local coordinate frame}: any vector $v\in\Reals^3$ that lies on triangle $\tri$ is a \textit{tangent vector}, which can be written equivalently as a 2-vector, $\tilde v \in\Reals^2$ in the local coordinate system of $\B_\tri$, as the unique vector satisfying $\sum_i \tilde v_i U_i=v$. 
While we derive intuition from the geometric 2-d tangent plane, we will alternatively treat it as the complex plane $\mathbb{C}$, with each tangent vector defined as $v\in\mathbb{C}$.

\paragraph{Piecewise linear functions and their gradients.} Hence, we follow the standard definition of piecewise linear functions (i.e. linear finite elements): 
we consider functions that are scalars assigned to the mesh's vertices. We denote such a function as $s\in\Reals^{|\V|}$, with $s_i$ being the scalar value associated with vertex $i$. Such scalar values on the vertices of a single triangle, $s_i,s_j,s_k$ uniquely define an affine function $a(p):\tri\to\Reals$ over the triangle, which interpolates these three values at the triangle's vertices, i.e., $a(v_i)=s_i$. Thus, the signal $s$ defines a piecewise-linear function over the triangles of the mesh, i.e., it is a linear function when restricted to one of the triangles. Therefore, its gradient is constant within the triangle and is a tangent vector that we denote (in the local coordinate system $\B_\tri$) as $f\in\mathbb{R}^2$. The gradient $f_\tri$ of a specific triangle $\tri$ can be obtained from the vertex values by applying the linear gradient operator which we denote $\nabla_\tri$, i.e., $f_\tri = \nabla_\tri (s_i,s_j,s_k)$. We use the notation $f$ to refer to the tensor of stacked gradient vectors over all the faces of the mesh. Finally, we consider multiple simultaneous signals, called \emph{channels}, $s^1,s^2,...,s^c$, where each channel $s^i$ is a scalar field, whose corresponding gradient field is $f^i$.

\paragraph{Poisson's equation.} As with many graphics applications that operate in the gradient domain, our method relies on the variational formulation of Poisson's equation. Given a set of tangent vectors $f$, the variational perspective of the Poisson equation finds \emph{the scalar function whose gradient best matches $f$.} In the continuous setting, this translates to the following least squares variational objective
\begin{equation}
        u = \min_{s}\int_{\Omega}||\nabla s- f||^2 dA.
        \label{eq:poissonVariational}
\end{equation}
Since our domain is a \emph{mesh}, $f$ is constant over each face. Hence, the above integrand becomes constant over each triangle, making the least squares problem reduce to a sparse linear system (see Eq. \ref{eq:poissonDiscrete}).

\section{PoissonNet}
\label{sec:method}
\begin{figure}
    \centering
    \includegraphics[width=\linewidth, trim=0 0.15cm 0 0]{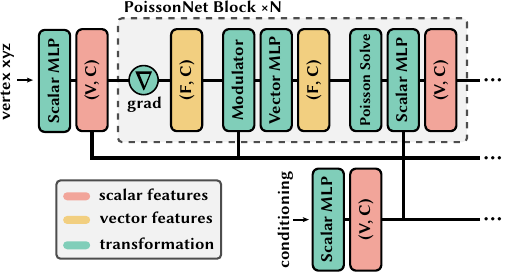}
    \caption{PoissonNet stacks identical blocks that transform features locally in the gradient domain and globally in the functional domain.}
    \label{fig:standardPipeline}
\end{figure}
Our method builds on the insight that using Poisson's equation as the core mechanism in an intrinsic learning architecture leads to several desirable properties. Poisson's equation can be solved \emph{efficiently} without resorting to lossy spectral approximations while serving as a truly global operator over the surface. These characteristics make PoissonNet a strong backbone for learning highly detailed signals on surfaces, where spectral approximations fail; learning deformations, where global operators are necessary; and common semantic tasks, such as mesh segmentation.
 
Our architecture is comprised of \emph{PoissonNet blocks} (depicted in Figures \ref{fig:pipeline} and \ref{fig:standardPipeline}), which interleave two simple operations on the incoming scalar features $s$:
\begin{enumerate}
    \item \emph{Local step in the gradient domain (Sec \ref{sec:method_gradientDomainLearning}).} Compute the gradient of the incoming signal, $\nabla s$, and locally (per-face) apply a learned transformation, producing gradient field $f$.
    \item \emph{Global step (Sec \ref{sec:method_PoissonSolve}).} Solve Poisson's equation using $f$, inducing a scalar update, $u$, that globally couples features on all vertices. Finally, output a learned transformation on $s$ and $u$.
\end{enumerate}
In the following sections we describe these steps in detail.

\subsection{Local step in the gradient domain}
\label{sec:method_gradientDomainLearning}
Given a scalar feature field with $C$ channels defined on the vertices of a mesh, $s \in \Reals^{\mathrm{|V| \times C}}$, we first compute its corresponding gradient field, $f \in \Complex^{\mathrm{|F| \times C}}$, using the intrinsic gradient operator
\begin{equation}
    f \coloneq \nabla s.
\end{equation}
These vector-valued quantities reside in the tangent space of each triangle, which is denoted by an orthonormal basis $\{u_1, u_2, n\}$ where $u_1$ and $u_2$ span the tangent plane and $n$ is the face normal. We express these gradients as a complex number $z = a+bi$ with $a$ and $b$ being the coefficients of $u_1$ and $u_2$, and hence transformations of these quantities can be represented by products with complex numbers, which induces a scaling and rotation within the tangent plane.

As is common when learning transformations of vector-valued quantities, we parametrize gradient transformations with learned complex weight matrices ~\cite{deltaconv,diffusionnet,vectorHeatNetwork,HSN_wiersma_2020}. Geometrically, the transformed gradient features become linear combinations of rotated and scaled gradients at each face. Notably, the choice of basis at each face is arbitrary up to a rotation (i.e. any orthonormal basis in the triangle can be chosen). Hence, it is desirable for our gradient transformation to maintain equivariance under linear coordinate transformations. As proposed by \citet{HSN_wiersma_2020}, we apply non-linearities to gradient magnitudes only, which preserves equivariance by ensuring that the gradient's directional component transforms consistently with the underlying coordinate system --- i.e. the phase of each gradient feature transforms linearly. The gradient transformation can be written succinctly on the $i$-th face as 
\begin{equation}
    \*f_i \leftarrow \*W\*f_i \odot \frac{\sigma(\*r + \*b)}{\*r}
\end{equation}
where $\*f_i \in \Complex^{\mathrm{C}}$ are the incoming gradient features on the face, and $\*W \in \Complex^{\mathrm{C \times C}}$ is a learned complex weight matrix. We write $\*r \in \Reals^{\mathrm{C}}$ as the vector holding magnitudes of each element of $\*W\*f_i$. Finally, $\*b \in \Reals^{\mathrm{C}}$ is a learned bias parameter, and $\sigma$ denotes a non-linearity; $\odot$ and division by $\*r$ are taken elementwise.

To enrich our gradient transformations, we additionally use the original scalar signal, $s$, to modulate the incoming gradients (before the transformation above). This allows the network to more discriminately transform gradient features using the scalar signal as intrinsic positional information. Let $s_\mathrm{face} \in \Reals^{\mathrm{F\times C}}$ denote the scalar features averaged onto faces. We modulate the phase and magnitude of gradient features via the element-wise product
\begin{equation}
    f \leftarrow (\sigma(\boldsymbol\gamma) + \epsilon) f e^{i\boldsymbol\theta}\quad \mathrm{where}\;\,\boldsymbol\gamma,\boldsymbol\theta = \mathrm{MLP}(s_\mathrm{face})
    \label{eq:modulation}
\end{equation}
where the scale factors $\boldsymbol\gamma$ and angular rotations $\boldsymbol\theta$ are given by a small multi-layer perceptron acting point-wise on $s_\mathrm{face}$. We apply a softplus activation, $\sigma$, to the scale factors; adding a small epsilon ensures positivity.

\subsection{Globally propagating gradient features}
\label{sec:method_PoissonSolve}
Once gradient-domain features have been transformed, we integrate them back into scalar features via a Poisson solve. Concretely, let $\*f \in \Reals^{\mathrm{2F \times C}}$ represent the stacking of components of all face-based transformed gradient features. We recover a global update to the scalar features by solving on each channel the sparse linear system
\begin{equation}
    \textbf{L} u = \nabla^\T\textbf{M}f,
    \label{eq:poissonDiscrete}
\end{equation}
where $\textbf{L}$ is the cotangent Laplacian~\cite{pinkall1993computing}, $\textbf{M}$ is the mesh's mass matrix, and $\nabla^\T$ represents the divergence operator. Since the solution $u$ is unique only up to an additive constant, we nullify its mean, centering the solution at zero. Divergence being a coordinate-free operator means that the Poisson solution is invariant to the choice of tangent bases. Finally, we apply a point-wise MLP to the concatenation of the input features $s$ and Poisson solution $u$. The feature update on the $i$-th vertex becomes
\begin{equation}
    s_i \leftarrow \mathrm{MLP}([s_i, u_i, c_i]).
    \label{eq:blockMLP}
\end{equation}
where $c_i$ are experiment-specific conditional features (see Sec. \ref{sec:results}).

Since the Poisson equation is an elliptic PDE, it can be solved efficiently without approximate timestepping or spectral methods, and its discretization admits a single pre-factorable sparse linear system that can be reused across all network blocks. These qualities allow PoissonNet to 1) efficiently scale, both in the size of datasets and the meshes themselves (i.e. the number of vertices); and 2) forgo lossy spectral approximations that are used in previous methods. 

\newcommand{\insetGreensFunction}{
\setlength{\intextsep}{0pt}
\setlength{\columnsep}{0.5em}
  \begin{wrapfigure}[6]{r}{55pt}
    \centering
    \includegraphics[width=\linewidth]{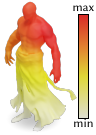}
  \end{wrapfigure}
}
\paragraph{Remark.}
Applying Poisson's equation to our network's gradient features is analogous to that of a global attention mechanism with fixed geometry-dependent weights. The inverse Laplacian $\*L^{-1}$ implicitly defines a Green’s function $G(i,j)$ that weights the contribution of the divergence at vertex $j$ to the update at vertex $i$. In this view, $G(i,j)$ serves as a global attention kernel over the mesh, aggregating gradient‑domain signals from all triangles\insetGreensFunction{}into each vertex’s scalar feature update---the inset visualizes $G$ w.r.t a vertex on the character's left shoulder. This construction is efficient, in that the attention kernel is defined through sparse mesh operators (rather than materializing a quadratic attention matrix), and has the added benefit of naturally adapting to the underlying mesh geometry.

%% file: sec_implementation.tex
\section{Implementation}
\label{sec:implementation}
\newcommand{\insetCUDAtiming}{
\setlength{\intextsep}{0pt}
\setlength{\columnsep}{0.8em}
  \begin{wrapfigure}[8]{r}{90pt}
    \centering
    \includegraphics[width=\linewidth]{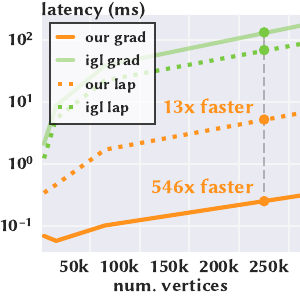}
  \end{wrapfigure}
}
\paragraph{Efficient construction of operators.}
We employ a custom PyTorch CUDA extension for the construction of our discrete mesh operators; namely the Laplacian, gradient, and mass matrices, allowing our training pipeline to forgo lengthy precomputation and instead compute necessary operators efficiently on-the-fly during training. This greatly reduces friction in experimentation and allows our method to be applied directly to large datasets. Notably, our method does not require precomputing a Laplacian eigenbasis, which often relies on CPU-based generalized eigendecomposition routines that are too slow to use during training and may take several hours to precompute even for moderately sized datasets. These qualities make PoissonNet more practical for large scale training and rapid experimentation, and additionally more flexible in pipelines\insetCUDAtiming{}with non-static training examples (e.g. when applying data augmentation to meshes). The inset compares our CUDA kernels against LibIGL \cite{libigl}. Our CUDA kernels emit sparse mesh operators directly using PyTorch's COO representation and support batching for homogeneous meshes (i.e. those with identical connectivity structure).

\paragraph{Solving Poisson's equation.}
We discretize the Poisson equation as in Equation \ref{eq:poissonDiscrete}. Our Poisson systems are solved using a shared Cholesky factorization of the Laplacian matrix, $\*L$, across all network blocks and channels, and hence the simultaneous per-channel linear systems are efficiently solved in parallel. We use Cholespy \cite{largeStepsInvRendering}, a CUDA-based sparse linear solver. Our Laplacian uses zero Neumann boundary conditions. Following Poisson’s variational form (Eq. \ref{eq:poissonVariational}), the inhomogeneous Neumann condition, $\partial u / \partial n = f\cdot n$, appears naturally, with $n$ being the outward boundary normal. The Poisson solution, $u$, is centered to obtain a unique solution.

%% file: sec_results.tex
\begin{figure}[ht]
    \centering
    \includegraphics[width=\linewidth, trim=0 0.25cm 0 0]{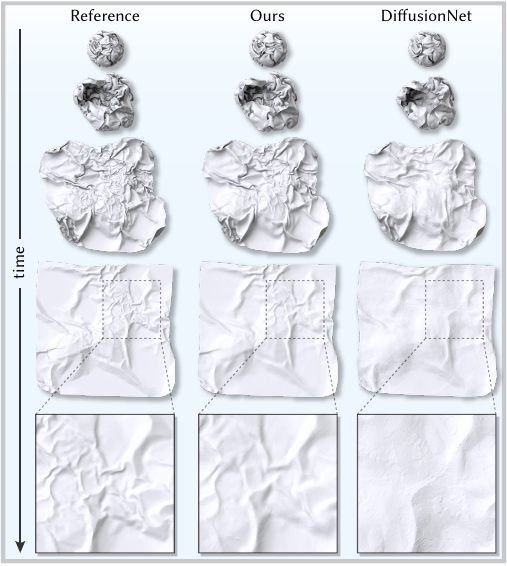}
    \caption{Comparison of architectures for representing highly-detailed signals. The networks are used to parametrize the evolution of a crumpling paper through time w.r.t. a rest configuration (see supplemental video). The mesh has 300k triangles and is available on TurboSquid \cite{crumplingPaperBall}.}
    \label{fig:crumpleComparison}
    \vspace{-2em}
\end{figure}
\begin{figure*}[t]
    \centering
    \includegraphics[width=\linewidth, trim=0 0.25cm 0 0]{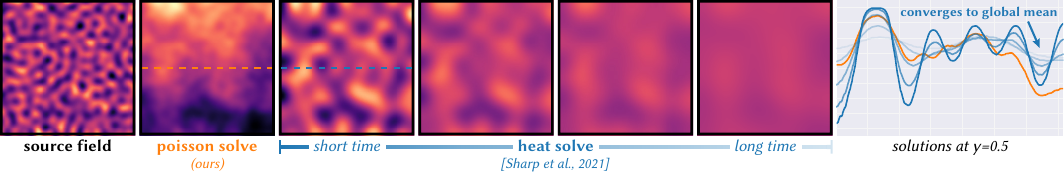}
    \caption{Poisson and heat solutions on an input scalar field. The Poisson equation (inhomogeneous) retains structure; whereas the heat equation (homogeneous) converges to a global mean, losing all structure over long-time heat flow. \textit{Right plot}: solution values along a cross-section at $y=0.5$ (dotted lines).}
    \label{fig:homogeneous}
\end{figure*}

\section{Results and Experimentation}
\label{sec:results}
In the following section, we evaluate PoissonNet on several applications, comparing it to current state‑of‑the‑art in learning on meshes. We focus on methods that perform intrinsic learning using differential operators, as the limitations of previous approaches have been demonstrated. See Section~\ref{sec:relatedWork} for a full discussion of these methods.

\paragraph{Experimental setup.} Across experiments we employ the same PoissonNet with varying numbers of network blocks, using 128-width blocks in all experiments. We use \texttt{xyz} vertex coordinates as our input features unless otherwise specified. Data augmentation is applied when applicable; we specifically augment shape orientation and global scale. Our experiments primarily compare to DiffusionNet \cite{diffusionnet} and DeltaConv \cite{deltaconv}, as they are leading methods for learning on surfaces using differential operators. We include further details in Section \ref{supp:exps}.

\newcommand{\insetPowerSpectrum}{
\setlength{\intextsep}{0pt}
\setlength{\columnsep}{0.5em}
  \begin{wrapfigure}[14]{r}{90pt}
    \centering
    \includegraphics[width=\linewidth]{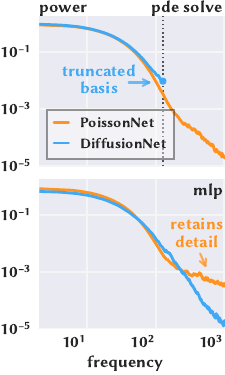}
  \end{wrapfigure}
}
\newcommand{\insetCrumpleLoss}{
\setlength{\intextsep}{0pt}
\setlength{\columnsep}{0.5em}
  \begin{wrapfigure}[8]{r}{90pt}
    \centering
    \includegraphics[width=\linewidth]{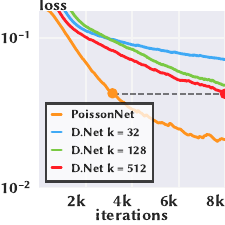}
  \end{wrapfigure}
}
\newcommand{\insetSpectrumLossCombined}{
\setlength{\intextsep}{0pt}
\setlength{\columnsep}{0.5em}
  \begin{wrapfigure}[23]{r}{90pt}
    \centering
    \includegraphics[width=\linewidth]{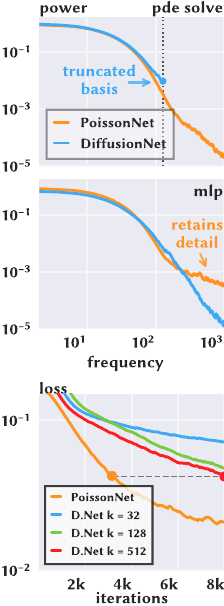}
  \end{wrapfigure}
}
\subsection{Analysis of Full-Spectrum Learning}
\label{sec:crumpleExp}
\insetPowerSpectrum{}Figure \ref{fig:crumpleComparison} demonstrates that our method is capable of representing extremely rich geometric signals. Here, we train PoissonNet to represent the evolution of an animated crumpling paper ball that has 300k faces \cite{crumplingPaperBall}. We parametrize the sequence by a scalar time $t$, which is used to condition the input of each network block's MLP (as in Eq. \ref{eq:blockMLP}). To further challenge our method, we only use a total of ${\sim}650k$ parameters, whose memory footprint (i.e., compression ratio) constitutes $2\%$ percent of the original sequence size; nevertheless, our method manages to preserve most of the fine details of the geometry.

We compare our method's performance to that of DiffusionNet~\cite{diffusionnet} by training it with the same number of parameters. Due to the limitations of the heat equation discussed in Section \ref{sec:relatedWork} and Figure \ref{fig:homogeneous}, DiffusionNet exhibits clear loss of detail and over-smoothing, struggling to represent the high-frequency wrinkles of the crumpled paper. The first inset further compares power spectra of the feature maps learned by both networks, confirming that our network is able to use higher frequency features to\insetCrumpleLoss{}represent the evolving geometry, while DiffusionNet encounters the expected issues that arise from the use of the heat equation. Additionally, the inset loss plot shows the clear effect of DiffusionNet's eigenbasis size (denoted by $k$) on training dynamics, as compared to our method. Both methods employ the \emph{NJF} head described in Section \ref{sec:shapeReposing}.

\newcommand{\insetReposeLoss}{
\setlength{\intextsep}{0pt}
\setlength{\columnsep}{0.5em}
  \begin{wrapfigure}[8]{r}{90pt}
    \centering
    \includegraphics[width=\linewidth]{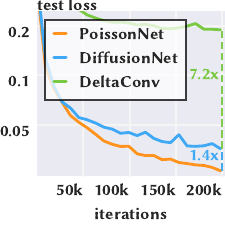}
  \end{wrapfigure}
}
\newcommand{\insetDeltaConvOutputs}{
\setlength{\intextsep}{0pt}
\setlength{\columnsep}{0.0em}
  \begin{wrapfigure}[4]{r}{40pt}
    \centering
    \includegraphics[width=\linewidth]{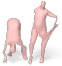}
  \end{wrapfigure}
}

\subsection{Shape Deformation}
\label{sec:shapeReposing}
We demonstrate that PoissonNet is capable of accurate, global reasoning, by learning to repose arbitrary humanoid character models without canonical poses or rigs, which requires global understanding of input geometry (joint articulation is an inherently long-range phenomenon, acting on kinematic chains). In particular, we accrue a dataset of 16k source-target mesh pairs generated by the SMPL-X human body model, using poses from the MOYO dataset~\cite{moyo_tripathi_2023, SMPL-X_pavlakos_2019}. These poses are comprised of motion-captured yoga sequences---containing ``pretzel-like'' contortions of human bodies, which are significantly more challenging to repose than traditional body poses. We deform a given source mesh into the target pose, conditioning the network on the SMPL-X pose parameters of the target. Our network uses five PoissonNet blocks, totaling 1.4 million parameters.

To conduct a fair comparison between network backbones, we employ the \emph{NJF} deformation head proposed by \citet{NJF}, which is a state-of-the-art method for parametrizing deformations. Briefly, the \emph{NJF} head receives as input three gradient fields associated with the gradients of the deformation map's $x,y,z$ components---i.e., a per-face Jacobian, $J_i\in\Reals^{2\times3}$. The \emph{NJF} head then produces a final deformation by solving Poisson's equation w.r.t the predicted Jacobians. We modify each architecture to predict the necessary Jacobian field (see Section \ref{supp:shapeDeformation} for details), and supervise the predicted deformations using \emph{NJF}'s proposed loss,
\begin{equation}
    \mathcal{L}_{\mathrm{NJF}} = \sum m_i \cdot\|v_i^{\mathrm{tar}} - u_i\|^2 + \sum \alpha_i \cdot \|J_i^{\mathrm{tar}} - \nabla u_i\|^2,
\label{eq:njfLoss}
\end{equation}
where $m$, $\alpha$ hold lumped vertex masses and face areas, and
$u$ is the solution to Equation \ref{eq:poissonDiscrete}. The ground truth vertex positions and Jacobians are denoted as $v^\mathrm{tar}$ and $J^\mathrm{tar}:=\nabla v^\mathrm{tar}$ respectively. 

\begin{figure*}[p]
    \centering
    \includegraphics[width=1.0\textwidth]{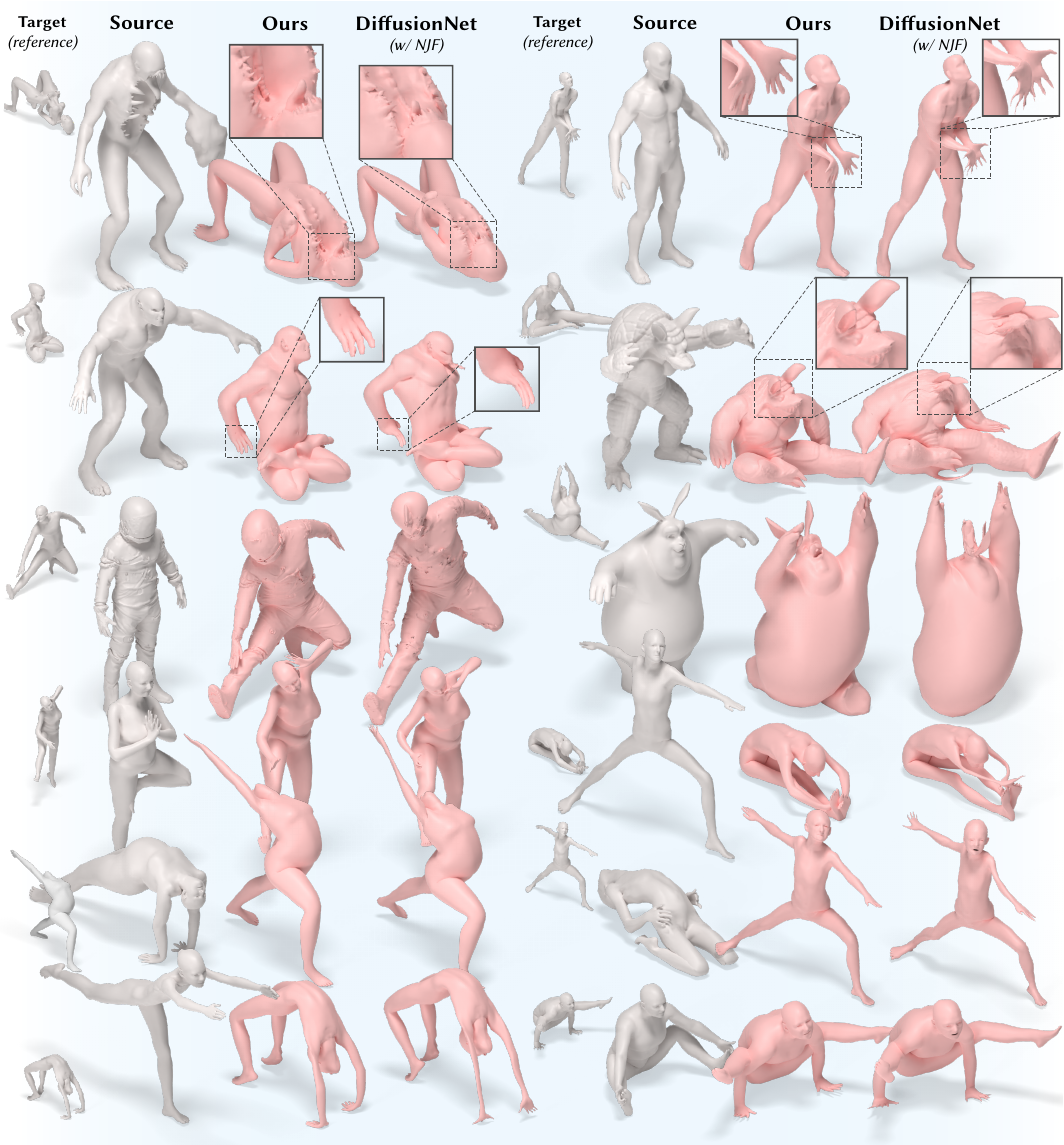}
    \vspace{-1em}
    \caption{
    We demonstrate that PoissonNet surpasses state-of-the-art intrinsic learning backbones for representing source-to-target deformations of humanoid characters. Here, we compare our deformations to that of DiffusionNet \cite{diffusionnet}, using diverse poses from the MOYO dataset \cite{moyo_tripathi_2023}. PoissonNet is able to faithfully match target poses while retaining intricate surface details from the source geometry, even for models that are substantially out of distribution (e.g. the top left mutant, and the Buck Bunny). We observe that DiffusionNet is unable to retain surface details, causing many body parts to become distorted (see close-ups). We additionally show in-distribution examples in the bottom half of the figure---of which these distortions remain noticeable.}
    \label{fig:fullPageYoga}
\end{figure*}

\begin{figure*}
    \centering
    \includegraphics[width=1.0\textwidth]{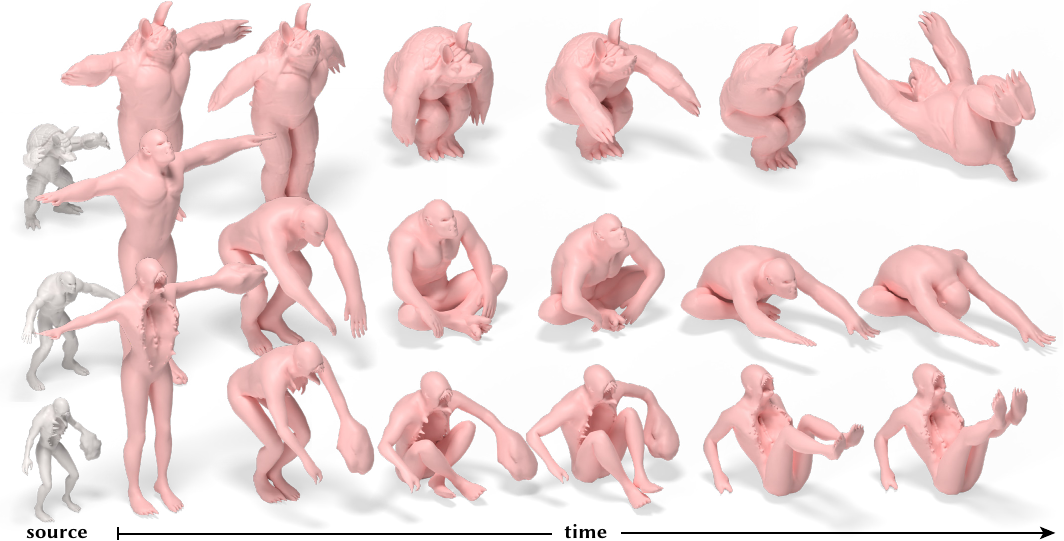}
    \caption{Our reposing network learns a smooth parametrization of humanoid poses that generalizes to out-of-distribution shapes. We repose several characters using pose sequences from the MOYO dataset \cite{moyo_tripathi_2023} and observe smoothly varying, realistic movements in accordance with that of a human.}
    \label{fig:sequentialYoga}
\end{figure*}

In \cref{fig:teaser,fig:meshyYoga} we show qualitative results of our method, and Figure \ref{fig:fullPageYoga} compares PoissonNet to that of a DiffusionNet backbone. Our network not only faithfully captures deformations of SMPL-X human bodies, but also boasts remarkable generalization to in-the-wild character models. We find that DiffusionNet is unable to retain high-frequency details in these shapes, often distorting their hands and faces (see Figure \ref{fig:fullPageYoga}). The inset plot reflects a similar conclusion: our model converges more quickly and reaches a much lower loss than the alternative backbones. We note that DeltaConv was unable to converge to a meaningful result on this\insetReposeLoss{}benchmark, likely due to its KNN-based differential operators, which are unreliable for surfaces with nearly-touching parts (e.g. in yoga poses). Finally, we show that PoissonNet is even able to transfer motion capture sequences to out-of-distribution characters, generating realistic motion sequences (see Figure \ref{fig:sequentialYoga} and supplemental video).
\newcommand{\insetYogaSegmentationAcc}{
\setlength{\intextsep}{0pt}
\setlength{\columnsep}{0.5em}
  \begin{wrapfigure}[9]{r}{90pt}
    \centering
    \includegraphics[width=\linewidth]{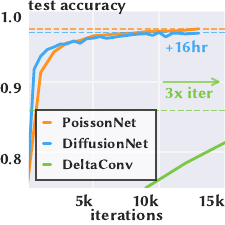}
  \end{wrapfigure}
}
\subsection{Semantic Segmentation}
\label{sec:segmentation}
We demonstrate that PoissonNet surpasses state-of-the-art performance on semantic segmentation of meshes, while remaining far more efficient than previous methods. In particular, we train a 3-block PoissonNet (650K parameters) on segmentations of the yoga motion capture shapes used in Section \ref{sec:shapeReposing} (totaling 32k training samples). We use the canonical SMPL-X segmentation map to delimit 27 unique body parts, distinguishing symmetric body parts (e.g. \emph{left}/\emph{right} forearm are separate classes). Our network achieves $97.03\%$ test accuracy; DiffusionNet achieves $96.12\%$ while requiring an additional \textit{16 hours of precomputation and 160gb of memory overhead due to the need for eigenbases}; and DeltaConv attains $88.2\%$ but is\insetYogaSegmentationAcc{}unable to reliably distinguish between left/right-sided parts, likely due to its local construction (see inset accuracy plot). Each method shows negligible variability in peak accuracy between runs ($<\!0.5\%$). We additionally train our network on the human body dataset of \citet{toriccover_maron_2017}. This dataset is an amalgamation of human meshes obtained from various sources \cite{faust2014,scape2005,adobe_2016,mit2008,giorgi2007shape}. The meshes are segmented into eight unique body parts. We report our test-time accuracy in Table \ref{table:segmentation1} alongside many previous methods as they were reported by \citet{deltaconv}. Our network matches state-of-the-art performance on this benchmark. Predicted segmentation maps for both datasets are shown in Figures \ref{fig:teaser} and \ref{fig:maronSeg}.

\newcommand{\insetEfficiencyPlot}{
\setlength{\intextsep}{0pt}
\setlength{\columnsep}{0.5em}
  \begin{wrapfigure}[17]{r}{90pt}
    \centering
    \includegraphics[width=\linewidth]{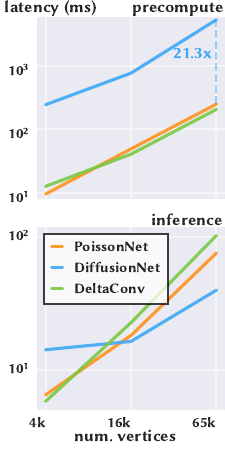}
  \end{wrapfigure}
}
\subsection{Classification}
We train PoissonNet on the SHREC11 shape classification benchmark \cite{shrec11dataset}, which contains 30 categories of shapes with 20 shape variations in each category. We employ a 3-block PoissonNet, identical to that of Section \ref{sec:segmentation}. Following previous methods \cite{meshcnn_hanocka_2019, deltaconv, diffusionnet,GWCNN_ezuz}, we train and test on simplified meshes, using just 10 examples per class for training, and averaging peak test accuracy over five training runs. Our network achieves a perfect accuracy of 100\% on the held out samples. Results are summarized in Table \ref{tab:shrec11}.

\begin{table}[]
\centering
\caption{Comparison of methods on SHREC11 shape classification.}
\label{tab:shrec11}
\rowcolors{2}{seabornblue!25}{seabornblue!75}
\begin{tabular}{@{}lr@{}}
\toprule
\textbf{Method}     & \textbf{Accuracy} \\ \midrule
MeshCNN\cite{meshcnn_hanocka_2019}   & 91.0\%   \\
HSN\cite{HSN_wiersma_2020} & 96.1\% \\
MeshWalker\cite{meshwalker_lahav_2020} & 97.1\% \\
PD-MeshNet\cite{primaldualconvo_milano_2020} & 99.1\% \\
HodgeNet\cite{hodgenet_smirnov_2021} & 94.7\% \\
FC\cite{fieldconvo_mitchel_2021} & 99.2\% \\
DiffusionNet\cite{diffusionnet} & {99.5\%}    \\
DeltaConv\cite{deltaconv} & {99.6\%}    \\
\textbf{PoissonNet (ours)} & {100.0\%} \\
\bottomrule
\end{tabular}
\vspace*{-1em}
\end{table}
\begin{figure*}
    \centering
    \includegraphics[width=1.0\textwidth,trim=0 0.25cm 0 0]{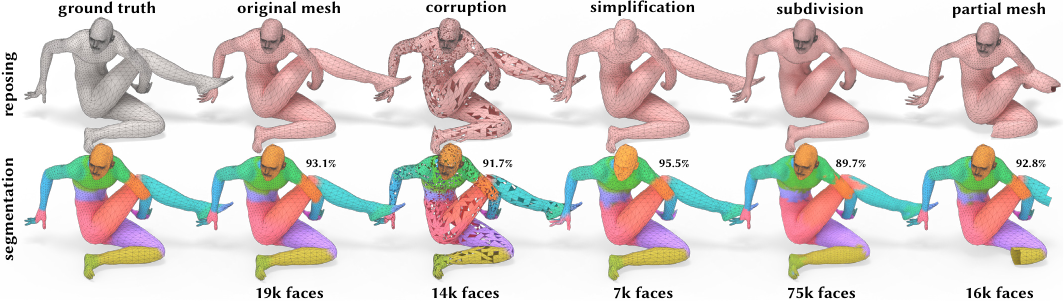}
    \caption{Our network is robust to changes in discretization. We apply our networks to a mesh under various such changes and observe stable predictions. From left to right: ground truths, our baseline prediction; and our predictions under corruption, quadric decimation, subdivision, and a partial (incomplete) mesh. For segmentations we provide accuracies w.r.t. the ground truth---in the case of topological changes, we use nearest-vertex matching to compute accuracy.}
    \label{fig:robustness}
\end{figure*}

\subsection{Analysis of Architectural Properties}
\paragraph{Training \& Inference Efficiency}
\insetEfficiencyPlot{}
PoissonNet's construction is efficient, making it straightforward to apply to large datasets and meshes with tens of thousands of vertices. Our method circumvents costly precomputation while being accurate and maintaining high throughput. These benefits extend to PoissonNet's forward latency on large meshes. In Table \ref{tab:efficiency}, we compare the training efficiency of our network with previous state-of-the-art methods on the experiment detailed in Section \ref{sec:shapeReposing}. Additionally, the inset figures compare the latency of these networks on meshes of increasing size. PoissonNet provides the best trade-off between precompute time, throughput, and accuracy. For fair comparison, we endow DiffusionNet with our CUDA kernels for precomputing Laplacian and gradient operators.

\input{tables/compute_efficiency}
We additionally compare the memory footprint of our Cholesky factorizations to methods that rely on Laplacian eigenbases. Although Cholesky factorization, in general, leads to possibly dense triangular matrices, we empirically find that our Cholesky factors scale within reasonable constant factors of $O(n\log n)$ nonzero elements, with $n$ being the number of mesh vertices and constant factors ranging from 3 to 6. For example, the Armadillo mesh subdivided to $n=800k$ vertices yields a decomposition requiring <200MB memory when accounting for sparse matrix storage overhead. A moderately-sized eigenbasis on the same mesh yields comparable memory usage; e.g., $k=128$ eigenvectors requires $n\cdot k \approx 390$MB.
\newcommand{\insetRobustGaps}{
  \begin{wrapfigure}[5]{r}[0pt]{90pt}
    \vspace{-\intextsep}
    \centering
    \includegraphics[width=\linewidth]{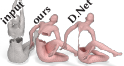}
  \end{wrapfigure}
}
\paragraph{Robustness.}
\insetRobustGaps
In Fig. \ref{fig:robustness} we show that PoissonNet's predictions remain stable under changes to triangulation; e.g. corruption, simplification, subdivision, and partial surfaces. Additionally, PoissonNet is more robust to surface holes compared to DiffusionNet, which introduces significant distortion around the hole (see inset). We attribute this to properties of Poisson's equation discussed by~\citet{naturalBoundaryCondOded}. Namely, the natural boundary conditions that appear in our inhomogeneous Poisson equation indicate that our feature fields will not become overly distorted around surface holes.

\begin{figure*}
    \centering
    \includegraphics[width=1.0\textwidth,trim=0 0.25cm 0 0]{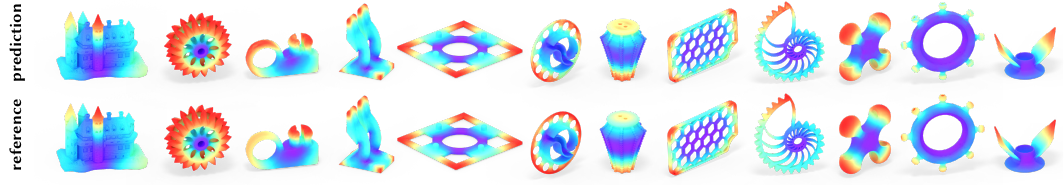}
    \caption{Outputs from our PoissonNet trained on multi-scale heat kernel signatures. Despite using a purely global operator, PoissonNet is able to capture HKS fields across a diverse array of shapes. All shapes shown are from the held out test set.}
    \label{fig:hksPrediction}
\end{figure*}
\paragraph{Learning local signals.}
Given that PoissonNet's spatial filter is a purely global operator, one may wonder if the network is suitable for learning local, multi-scale signals; e.g., could PoissonNet predict the heat kernel signatures (HKS) \cite{heatKernelSignature} on a given shape? We indeed find that our network is capable of learning such signals by training a 3-block PoissonNet on heat kernel signatures computed on shapes from the Thingi10k dataset \cite{Thingi10K} that have been pre-processed by fTetWild \cite{tetwild}. We retain all shapes that contain a single connected component and have more than 500 vertices, leaving 7500 shapes. We use an 85-15 train-test split. For each shape, we sample 16 HKS feature fields using time values logarithmically spaced in the interval $[0.01, 1]$. Each channel is independently normalized to $[0,1]$ to stabilize learning; hence, our HKS predictions are meaningful up to a global scaling. Our network predicts all feature maps simultaneously and is supervised using a standard MSE loss. Figure \ref{fig:hksPrediction} shows that PoissonNet is able to represent heat kernel signatures on a diverse array of out-of-distribution shapes.

\begin{figure}
    \centering
    \includegraphics[width=\linewidth, trim=0 0.25cm 0 0]{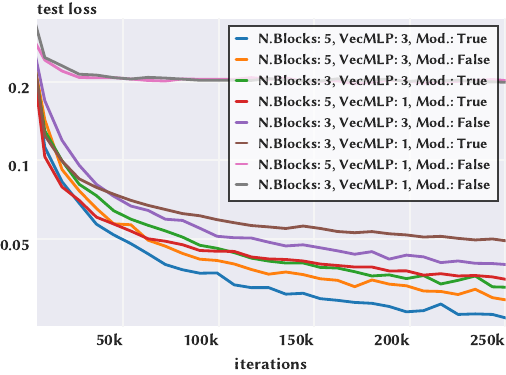}
    \caption{Hyperparameter sweep over various PoissonNet configurations for our shape deformation experiment. \emph{N.Blocks} denotes number of PoissonBlocks, \emph{VecMLP} denotes the number of layers in each block's Vector MLP, and \emph{Mod} denotes the usage of our proposed vector feature modulation step given by Equation \ref{eq:modulation}, which is applied before each Vector MLP.}
    \label{supp:ablationLoss}
\end{figure}
\paragraph{Ablating design choices.} Figure \ref{supp:ablationLoss} shows the effect of various design decisions with respect to the shape deformation experiment in Section \ref{sec:shapeReposing}. We observe that, all else being equal, employing our modulation step (Eq. \ref{eq:modulation}) improves network performance; similarly, using larger Vector MLPs (more layers) and more PoissonNet blocks steadily improves performance. The blue curve represents our primary network used in all results.

%% file: tables/compute_efficiency.tex
\begin{table}
    \centering
    \caption{Compute efficiency on our reposing experiment, alongside single-mesh forward latency. Our method has minimal precompute while maintaining high throughput and accuracy. By comparison, \textit{DiffusionNet (spectral)}’s eigenbasis precompute takes several hours and demands large memory overhead, while \textit{DiffusionNet (direct)} is too expensive to train. 
    For fair comparison, both PoissonNet and DiffusionNet use our operator CUDA kernels.}
    \label{tab:combinedVertical}
    \rowcolors{2}{seabornblue!75}{seabornblue!25}
    \begin{tabular}{l | c c c}
        \toprule
        \multicolumn{4}{c}{\textbf{Total compute expenditure}} \\
        \midrule
        \rowcolor{white}
        Method 
            & \thead{Precompute} 
            & \thead{Train}
            & \thead{Total} \\
        \midrule
        \textbf{PoissonNet}
            & $<\!1\,\mathrm{min}$ & 9058 batch/hr & 22.1hr  \\
        \makecell[l]{DiffusionNet\\\;\,\textit{(spectral)}}
            & \makecell[c]{+9hr\\+80GB mem.} & 11438 batch/hr & 26.5hr \\
        DeltaConv
            & $<\!1\,\mathrm{min}$ & 9015 batch/hr & 22.2hr \\
        \midrule
        \addlinespace[0.25em]
        \rowcolor{white}
        \multicolumn{4}{c}{\textbf{Single mesh forward latency} (\textit{precompute}+\textit{inference} in ms)} \\
        \midrule
        \rowcolor{white}
        Method 
            & \thead{4k verts} 
            & \thead{16k verts} 
            & \thead{65k verts} \\
        \midrule
        \rowcolor{seabornblue!75}
        \textbf{PoissonNet}
            & 9.6+6.5ms & 49.4+18.2ms & 251+75.9ms \\
        \rowcolor{seabornblue!25}
        \makecell[l]{DiffusionNet\\\;\,\textit{(spectral)}}
         & 245+14.2ms & 770+16.4ms & 5340+39.9ms \\
        \rowcolor{seabornblue!75}
        DeltaConv
            & 12.7+5.8ms & 40.8+22.7ms & 207+103ms \\
        \bottomrule
    \end{tabular}
    \label{tab:efficiency}
\end{table}

%% file: sec_conclusion.tex
\section{Conclusion}
We have demonstrated that PoissonNet's local-global approach sidesteps several trade-offs associated with previous intrinsic approaches --- resulting in a network that is efficient, scalable, and robust to out-of-distribution geometry. It serves as a practical tool enabling several learning applications, including animation of intricately detailed character models (without rigs), semantic segmentation, and compression of high-frequency geometry. 

\emph{Limitations.} 
Our method is not well suited for learning on shapes with multiple connected component; as our intrinsic Poisson equation will operate indendently on each component. This drawback is also shared by other intrinsic methods based on differential operators~\cite{diffusionnet,hodgenet_smirnov_2021,HSN_wiersma_2020}. The utility of these networks would be greatly expanded if they generalized to multi-component meshes, which dominate the corpus of in-the-wild models. Facilitating this learning in a principled way, without ad-hoc KNN-based solutions, remains an unsolved problem. Moreover, extremely poor discretizations (e.g. sliver triangles) may lead to numerical instability for each of these approaches; e.g., by distorting the solution to Poisson's equation. Next, although PoissonNet requires less computational resources than comparable methods, it is still too slow to enable real-time applications on very large meshes. Finally, while each block of PoissonNet has global support, it has a decaying receptive field; i.e., the coupling between vertices diminishes with distance. We consider the investigation of more general classes of signal propagation using different PDEs, such as the \emph{wave equation}, as important future work.
\begin{figure}[b]
    \centering
    \includegraphics[width=\linewidth, trim=0 0 0 0]{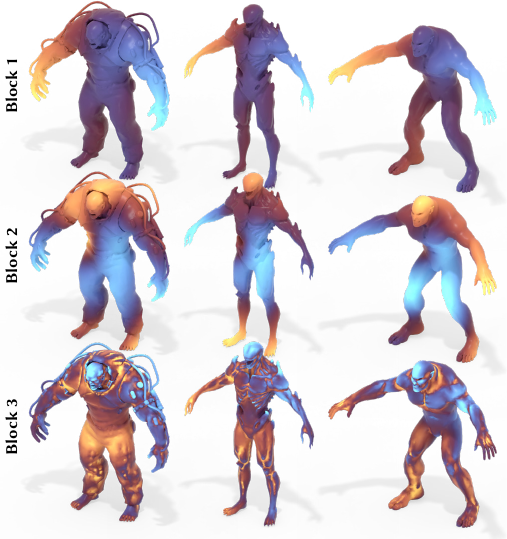}
    \caption{Feature maps given by the Poisson solves across a pre-trained three block PoissonNet. We visualize a fixed channel for all feature maps and find their qualitative appearance to be similar across shapes.}
    \label{fig:featureMaps}
\end{figure}

%% file: supp.tex
\section{Experimental and Implementation Details}
\label{supp:exps}
\begin{figure*}[h!]
    \centering
    \includegraphics[width=0.9\textwidth,trim=0 0.5cm 0 0]{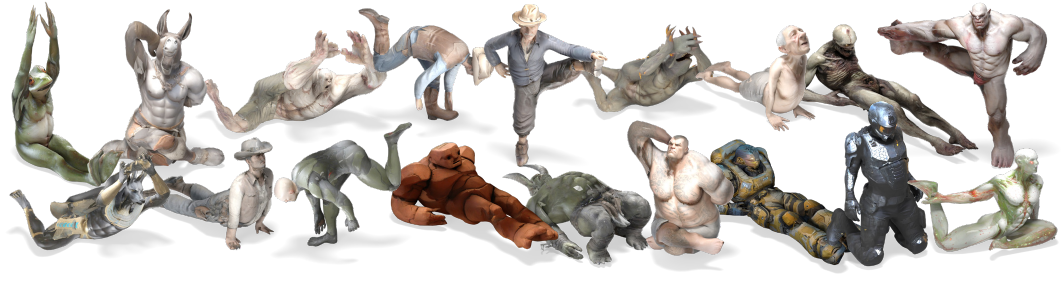}
    \caption{We further demonstrate that PoissonNet can be used to repose characters that are produced by generative models. Here, we show a diverse selection of characters generated by \citet{MeshyAI} that have been reposed using PoissonNet with pose inputs from the MOYO dataset \cite{moyo_tripathi_2023}.}
    \label{fig:meshyYoga}
\end{figure*}
\begin{figure}
    \centering
    \includegraphics[width=0.95\linewidth]{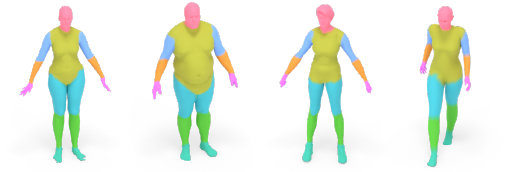}
    \caption{PoissonNet's segmentations on the dataset of \citet{toriccover_maron_2017}.}
    \label{fig:maronSeg}
\end{figure}

\begin{table}[b]
    \caption{Comparison of methods on the human mesh segmentation task of \citet{toriccover_maron_2017}. Table is provided by \citet{deltaconv}.}
    \label{table:segmentation1}
    \begin{center}
        \rowcolors{2}{seabornblue!25}{seabornblue!75}
        \begin{tabular}{lr}
            \toprule
            Method     & Accuracy \\
            \midrule
            PointNet++~\cite{pointnet++_qi_2017} & 90.8 \\
            MDGCNN~\cite{mdgcnn_poulenard_2018} & 88.6 \\
            DGCNN~\cite{dgcnn_wang_2019} & 89.7 \\
            SNGC~\cite{sngc_haim_2019} & 91.0 \\
            HSN~\cite{HSN_wiersma_2020}  & 91.1 \\
            MeshWalker~\cite{meshwalker_lahav_2020} & \textbf{92.7}\\
            CGConv~\cite{cgconv_Yang_2021}  & 89.9 \\
            FC~\cite{fieldconvo_mitchel_2021} & 92.5 \\
            DiffusionNet - xyz~\cite{diffusionnet} & 90.6    \\
            DiffusionNet - hks~\cite{diffusionnet} & 91.7     \\
            DeltaConv~\cite{deltaconv} & 92.2 \\
            \midrule
            PoissonNet - xyz & 90.7 \\
            PoissonNet - hks & 91.1 \\
            \bottomrule
        \end{tabular}
        % }
    \end{center}
\end{table}

Below we provide details for all experiments outlined in the main text, including the employed PoissonNet architectures, details of compared methods, and training hyperparameters.

\subsection{Analysis of Full-Spectrum Learning}
We train a 128-width PoissonNet with two blocks using an additional \emph{NJF} head. Each block uses a VectorMLP of two layers. We train our network on all frames of the crumpling paper ball dataset (see Sec. \ref{supp:paperBallDataset}) using a learning rate of 0.001 and virtual batch size of 8. The employed DiffusionNet baseline uses similar parameters; however, for fair comparison we endow DiffusionNet with wider blocks, using a width of 192 to equalize the total number of parameters, and an \emph{NJF} head is provided. As with our shape deformation experiment, we validated the baseline with and without an NJF head---in this particular experiment, we found the results to be moderately better with the head included. The paper crumpling sequence is parametrized by a single scalar time value, which is appended to the scalar MLPs of both networks.

\paragraph{Analysis of network features.} In Section \ref{sec:crumpleExp} we compare the power spectra of PoissonNet's learned features compared to that of DiffusionNet. In particular, we extract feature maps from the last network block (before the appended \emph{NJF} heads) for both networks, and project these features into an eigenbasis given by the crumpled paper's Laplace-Beltrami operator. The eigenbasis has size $K=1024$. Let $\*f\in\Reals^{K \times C}$ denote the projection of a feature map with $C$ channels into the spectral basis. We compute the power spectra per-channel $c$, $p^c_k$, over all $\*f_k$ with $k\in[1,K]$ via, $p^c_k = |\*f^c_k|^2$,
normalizing the power channel-wise, $\hat{p}^c_k = p^c /\sum^K_j p_j^c$. Finally, we compute the maximum channel-wise power, taking that as our final power spectrum. The first inset in Section \ref{sec:crumpleExp} shows power spectra for feature maps produced by the networks' PDEs and MLP layers respectively. Our method produces features with more power in the high-frequency range, notably our MLP's features retain higher frequencies in the band that is truncated by DiffusionNet's spectral PDE solve.

\subsection{Shape Deformation}
\label{supp:shapeDeformation}
In our shape deformation experiments we use a 128-width PoissonNet with five blocks, using VectorMLPs with three layers. We train using a learning rate of 0.0005 and batch size of 16. The employed DiffusionNet baseline uses equal number of blocks with a larger width of 192 to equalize parameter count. We experimented with various configurations for DeltaConv, which we discuss in a later paragraph.
Finally, all networks are endowed with an additional \emph{NJF} head to parametrize the deformation---we experiment with alternative choices, discussed in paragraph \emph{NJF Head} below.

\paragraph{Experimental setup.}
We employ our full dataset of 16k source-target SMPL-X human body pairs with poses sourced from the MOYO dataset \cite{moyo_tripathi_2023,SMPL-X_pavlakos_2019}---see Section \ref{supp:MOYODataset} for details. Each network is conditioned on all 153 SMPL-X pose parameters---parameters for eyes are excluded. We experimented with two choices for injection of these parameters into the tested backbones: 1) by simply included these parameters in the input layer of the networks, 2) by concatentation of the parameters in each network block's MLP. We found that the second option is superior for all backbones. We employ standard data augmentation techniques to further improve generalization; in particular, we apply random shifts and scalings to the training shapes. In the case of DiffusionNet, which uses precomputed eigenbases, we are mindful to update them accordingly when scaling training shapes dynamically.

\paragraph{NJF Head.}
We experiment with two parametrizations of the deformation map: 1) a direct prediction of the target vertices (i.e. networks predict the deformed \emph{xyz} coordinates directly), and 2) appending an \emph{NJF} head to the end of each network. We find that using an \emph{NJF} head is superior across all backbones, especially for generalization to out-of-distribution geometries. We implement identical \emph{NJF} heads for each method. Concretely, the head expects input features on vertices, which are then mapped to corresponding gradient features on faces using the intrinsic gradient operator. We transform these features using a Vector MLP that maps the input gradient features into a 2-vector field for each of the target coordinate channels (i.e. \emph{x,y,z})---these serve as the predicted mapping's Jacobian fields, which are then integrated using Poisson's equation to produce the final deformed vertices. We additionally add the source shape's \emph{xyz} gradients to the \emph{NJF} head's predicted Jacobians, serving as a skip connection in the gradient domain. We re-iterate that identical constructions were tested across backbones in these experiments.

\paragraph{DeltaConv Baseline.}
We employed various configurations for the DeltaConv backbone used in Section \ref{sec:shapeReposing}, ranging from networks that matched our parameter count of ${\sim}1.5$ million, to deeper networks totaling ${\sim}2.5$ million parameters. We additionally, disabled the BatchNorm layers in DeltaConv's MLPs, as they hindered deformation performance. However, we found that all configurations were unable to converge to a comparable result as compared to our method and DiffusionNet. We attribute this to two properties inherent in DeltaConv's construction: 1) the use of point-based K-NN surface operators, which are extremely noisy for surfaces that have near-touching geometry. This is especially relevant in our case, as our dataset contains yoga poses, where ground truth geometry is often ``kissing'', or even intersecting; 2) DeltaConv's network blocks use \emph{local} mesh operators, whereas our method features fully global support in each network block. For this reason, we experimented with DeltaConv backbones that were comparatively deeper (i.e. to enlarge its effective receptive field). However, we note that doing so only marginally improved performance, and in some cases actually hindered performance due to over-parametrization. We note that the employed DeltaConv architecture did not use spatial downsampling or upsampling layers (as in a UNet-style network), as this architecture variant was not available in the open source code. Using spatial pooling may improve performance, as it would effectively broaden the network's receptive field; However, critical limitations would remain (i.e. sensitivity to sampling density).

\subsection{Semantic Segmentation on MOYO Dataset}
We train a 128-width PoissonNet with three blocks, using VectorMLPs with a single layer. We train using a learning rate of 0.001 and batch size of 16. The DiffusionNet baseline uses blocks of 176 width to equalize parameter count, and the DeltaConv baseline was configured similarly. The inset loss plot in Section \ref{sec:segmentation} shows that our method converges to a higher test accuracy, while being far more compute efficient than the compared methods. Namely, DiffusionNet required an additional 16 hours of precompute and 160gb of memory overhead to compute and store the dense eigenbases over the entire segmentation dataset. Meanwhile, DeltaConv required 3 times as many training iterations to converge to a lower final test accuracy. The total wallclock time of our method was only 2 hours.

\section{Ablating design decisions}
In Figure \ref{supp:ablationLoss} we show validation loss curves for various hyperparameter configurations on our shape deformation experiment. We specifically highlight that our vector feature modulation step (given by Eq. \ref{eq:modulation}) clearly improves performance on this benchmark. This gain in performance is representative across all of our experiments.

\section{Creation of Datasets}
\subsection{MOYO Dataset}
\label{supp:MOYODataset}
We employ the MOYO dataset \cite{moyo_tripathi_2023} in our shape deformation and semantic segmentation experiments (see Sections \ref{sec:segmentation} and \ref{sec:shapeReposing}). In particular, we generate a dataset of 32k training and 4k validation SMPL-X \cite{SMPL-X_pavlakos_2019} human bodies using yoga poses sampled from MOYO. Because MOYO contains temporal motion captures, there are many near-identical frames that are redundant for our experiments; hence, we employ a greedy farthest point sampling routine to select a subset of the poses that are most distinct. We then generate human bodies by randomly sampling SMPL-X body shapes---we aggressively sample body parameters using a Gaussian distribution with a standard deviation of 5. This provides body models with more diverse geometry, which aids in generalization. The dataset is generated with pairs of source-target models with identical body shapes. For our semantic segmentation experiment, we use the canonical SMPL-X vertex segmentation map. In particular, we use all 27 classes---we exclude the eyes/eyeball classes. We use all 32 thousands training examples in both experiments; for shape deformation this constitutes 16 thousand source-target pairs.

\subsection{Crumpling Paper Ball}
\label{supp:paperBallDataset}
We use the simulated crumpling paper ball published by \citet{crumplingPaperBall} to benchmark our network. The sequence is comprised of 118 frames, each with a canonical mesh topology of 300k faces.